\documentstyle[12pt]{report}
\def\lsim{\lower0.6ex\vbox{\hbox{$ \buildrel{\textstyle <}\over{\sim}\ $}}}
\def\rsim{\lower0.6ex\vbox{\hbox{$ \buildrel{\textstyle >}\over{\sim}\ $}}}
\def\hel3{$^3$He}
\def\hel4{$^4$He}
\def\li7{$^7$Li}

\font\eightrm=cmr8

\begin{document}
\noindent
\phantom{DESY 90--???}           \hfill OSU-TA-24/94\\
 \phantom{DESY 90--???} 	 \hfill   (December 29, 1994) \\
 \vspace{ 1.2 cm}
\begin{center}
\begin{Large}
\begin{bf}
THE X-RAY CLUSTER \\
BARYON CRISIS\\
\end{bf} \end{Large}
\end{center}
\begin{center}
\begin{large}
Gary Steigman\\
\end{large}
 \vspace{0.3cm}
Departments of Physics and Astronomy,
The Ohio State University\\
174 West 18th Avenue,
 Columbus, OH 43210 \\
\vspace{0.3cm}
and\\
\vspace{0.3cm}
\begin{large}
James E. Felten\\
 \vspace{0.3cm}
\end{large}
Code 685, NASA
Goddard Space Flight Center\\
Greenbelt, MD 20771\\

\end{center}
 \vspace{.5cm}
\centerline{\bf ABSTRACT}
\begin{quotation}
\noindent
Nucleosynthesis in the standard hot big bang cosmology offers a successful
account of
the production of the light nuclides during the early evolution of the
Universe.
Consistency among the predicted and observed abundances of D, $^3$He, \hel4
and \li7
leads to restrictive lower and upper bounds to the present density of
nucleons.  In
particular, the upper bound ensures that nucleons cannot account for more
than a
small fraction $(<0.06 h{^{-2} _{50}})$ of the mass in a critical density
(Einstein-de Sitter) Universe.  In contrast, x-ray observations of rich
clusters of
galaxies  suggest strongly that baryons (in hot gas) contribute a
significant fraction
of the total cluster mass $(\ge 0.2 h{^{-3/2} _{50}})$.   If, indeed,
clusters do
provide a ``fair" sample of the mass in the Universe, this ``crisis" forces
us to
consider other ways of mitigating it, including  the politically
incorrect possibility that
$\Omega < 1$. The options, including magnetic or turbulent pressure,
clumping, and
non-zero space curvature and/or  cosmological constant, are discussed.
\end{quotation}
\vspace{.1in}
\noindent To appear in {\it Proceedings of the St. Petersburg Gamow
Seminar} (Sept.
12-14, 1994), ed. A. M. Bykov \& R. A. Chevalier; {\it Sp. Sci. Rev.}, in
press
(Dordrecht: Kluwer).
\vfil
\newpage

\noindent
1. INTRODUCTION

The standard, hot big bang cosmology provides a successful model of an
expanding
Universe filled with radiation.  As    Gamow and his
collaborators Alpher and Herman realized, the Universe described by this
model would
have passed through an early, hot, dense epoch when nuclear reactions
transformed
neutrons and protons into the light nuclides deuterium, helium-3, helium-4
and
lithium-7 (e.g., Boesgaard \& Steigman 1985).

The primordial abundances predicted by Big Bang Nucleosynthesis (BBN) in the
standard model depend on only one adjustable parameter -- the nucleon
density at
BBN.  For relatively high nucleon density models (as measured by the
nucleon--to--photon ratio $\eta = n_N/n_{\gamma}$; $\eta _{10} \equiv
10^{10}
\eta $),
 nucleosynthesis begins early, when neutrons are relatively
abundant.  In this case, D and $^3$He are quickly burned to \hel4, and it
is easier to
bridge the gap at mass-5 and produce relatively large yields of mass-7.
Thus for
``high"
$\eta$ the primordial abundances of D and $^3$He are ``small'' while those
of \hel4 and
\li7 are ``large".  In contrast, for relatively low nucleon density models,
the start
of BBN is somewhat delayed, permitting some neutrons to decay and resulting
in a
somewhat less efficient burning of deuterium and helium-3 (as well as
lithium-7).
So, for ``low" $\eta$ the big bang yields of D and $^3$He (as well as \li7)
are
``large" while that of \hel4 is ``small". With four predicted abundances
(relative
to hydrogen) and only one adjustable parameter, the standard, hot big bang
cosmology
is a testable model.  As a function of $\eta$ the predicted BBN yields
range over some
10 orders of magnitude. So, too, do the ``observed" primordial abundances
as inferred
from a wide diversity of astronomical observations [e.g., Walker et al.
(``WSSOK") 1991;
for a recent overview, see Steigman 1994a]. For quite some time now it has
been known
that theory and data are roughly consistent (at the $\sim$ 2-sigma level)
provided
that the nucleon abundance lies in a very narrow range:  $2.8~ \lsim \eta
_{10}~
\lsim 4.0$ [WSSOK; or: $3.1 ~\lsim \eta _{10}~ \lsim 3.9$ (Steigman 1994a)].

It is, of course, not sufficient to find that a value (or narrow range of
values) of
$\eta$ exists such that BBN  predicts correctly the primordial abundances
of the light
nuclides (although any cosmological model must pass this test).  It is
necessary,
too, to see if the nucleon abundance $\eta$ inferred from processes in the
youth of
the Universe is consistent with that determined from observations in its
maturity.
Astronomical data exist on the dynamics of systems from galaxies to
clusters of
galaxies (and beyond), from which estimates of the universal mass density
may be
derived.  The BBN inferred nucleon mass density must be compared with those
estimates to further test the standard, hot big bang cosmology.

For convenience (as well as convention) we will use for our comparisons
the density
parameter $\Omega$, the ratio of the present mass density to the critical
mass
density $\rho _{crit}$.
$$
 \rho _{crit} = {3 H{^2 _0} \over 8 \pi G} \approx 2.6 h {^2 _{50}} ~{\rm
keV
cm} ^{-3} .
\eqno (1) $$
\vfil\eject

\noindent In (1), $G$ is Newton's constant and we have written the Hubble
constant as:
$H_0 = 50 h_{50}~ {\rm km~s ^{-1} Mpc} ^{-1}$. The nucleon--to--photon
ratio $\eta$
(e.g.,
from BBN) and the present photon (Cosmic Background Radiation $\equiv$  CBR)
temperature determine the present universal nucleon density. For $T_{CBR}
\approx
2.73$~K,
$$
 \Omega _{BBN} h{_{50} ^2} \approx 0.015 \eta _{10} .
   \eqno(2)$$
[For $T_{CBR} = 2.726 \pm 0.010$ ~(Mather et al.\ 1994), $\Omega _{BBN}
h{^2
_{50}} /
\eta _{10} = 0.0146 {^{+0.0002} _{-0.0001}}$.] Thus, for $2.8 ~\lsim \eta
_{10} ~\lsim
4.0$ (see Fig. 1),
$$
 0.04 ~\lsim \Omega _{BBN} h{^2 _{50}} ~\lsim 0.06.
\eqno (3)$$

We show this allowed range of $\Omega _{BBN}$ as a function of $H_0$ in
Figure 1.
For comparison, we show also an estimate of $\Omega _{LUM}$, the
contribution to
$\Omega$ by ``luminous" baryonic matter; i.e., baryonic matter within
optically
visible galaxies. We obtained this value by assuming, within the luminous
parts of
galaxies, a mean ratio of baryonic mass to blue luminosity $\langle M/L_B
\rangle =
7.5 h_{50}$, and dividing by the critical ratio needed to obtain $\Omega =
1$,
$(M/L_B)_{crit} = 750 h_{50}$, inferred from surveys of the large-scale
luminosity
density (Efstathiou, Ellis, \& Peterson 1988). This estimate $\Omega _{LUM}
= 10
^{-2}$ is generous. Some would argue that it is an upper bound.

Comparing $\Omega _{BBN}$ and $\Omega _{LUM}$, we see that as required for
consistency, BBN has provided evidence for at least as many nucleons as
observed
today $(\Omega _{BBN} \ge \Omega _{LUM}$ for $H_0 \le 100~ {\rm km~s^{-1}
Mpc}^{-
1}$).
Indeed, for $H_0 < 100$, the gap in Fig. 1 between $\Omega _{BBN}$ and
$\Omega
_{LUM}$ provides evidence that a significant fraction of all nucleons in the
Universe may be ``dark" (baryonic dark matter) -- at least in the sense of
not being
within the optically visible parts of galaxies.

This argument has been made before. We should offer two cautionary remarks:
(1) Our
generous chosen ratio $\langle M/L_B \rangle = 7.5 h_{50}$ [cf.\ the value
$3.2
h_{50}$ used by White et al.\ (1993)] comes from dynamical determinations
(as shown by
the
$H_0$-dependence) and therefore includes all gravitating mass within the
images
studied, not just baryonic mass. If some fraction of this mass were non-
baryonic, our
estimate for {\it baryonic} mass $\Omega _{LUM} = 10 ^{-2}$ would have to
be {\it
decreased}. (2) There are few dynamical measurements in the outer parts of
galaxies.
If large amounts of baryonic matter were present there, the number 10$^{-
2}$ might
have to be {\it increased}. We shall discuss this further elsewhere (Felten
\&
Steigman 1994).

We also show in Figure 1 a lower bound $(\Omega _{DYN} > 0.1$) to the {\it
total}
mass density inferred from gravitational dynamics on the scales of clusters
and
beyond. If $H_0$ is not very small (e.g., for $H_0 \ge 40$), we see that
$\Omega
_{BBN} < \Omega _{DYN}$, implying that the universal mass density contains,
and may
be dominated by, non-baryonic dark matter (``The Ultimate Copernican
Principle"!).

\vfil\eject

\noindent 2. CONSISTENCY OR CRISIS ?

The dramatic quantitative success of BBN  provides
strong support for the standard, hot big bang cosmology. At the same time,
however,
it requires that we proceed to a higher level of accuracy in testing the
consistency
of this model. As we move on to more stringent tests, recent observational
data have
revealed three potential crises looming on the horizon (Steigman 1994a).

\noindent (a) Is the Predicted $^4$He Yield Too High?

The lower bound to $\eta$ and, therefore, to the BBN predicted yield of
\hel4, is
derived (solely!) from a upper bound to primordial D and/or $^3$He derived
from
solar system data (Yang et al.\ 1984; Dearborn, Schramm, \& Steigman 1986;
WSSOK;
Steigman
\& Tosi 1992, 1994). For $\eta _{10} \ge 2.8$ (WSSOK) and the
standard
case of three light neutrino species $(N_{\nu} = 3$) and a neutron lifetime
$\ge$ 885
sec, the predicted BBN \hel4 mass fraction is ``large": $Y_{BBN} \ge
0.241$. In
contrast, the primordial abundance inferred from observations of low
metallicity,
extragalactic H\ {\eightrm II} regions (Pagel et al.\ 1992, Skillman et al.
1993) is
``small". The
recent analysis of this large data sample by Olive \& Steigman (1994)
concludes
that $Y_P = 0.232 \pm 0.003$; the uncertainty is the 1-$\sigma$ statistical
error in
the mean. In the absence of any systematic uncertainties, the 2-
$\sigma_{stat}$ upper
bound is inconsistent with the theoretical lower bound. However, this
potential crisis
may be avoided with allowance for a modest ($\sim 2\%$) systematic
uncertainty in the
observationally inferred $Y_P$. Sources of such uncertainties could be small
corrections for unseen neutral helium, for collisional excitation, for
corrections
due to radiation trapping in the presence of dust, or for errors in the
atomic
emissivities.
  Although Olive \& Steigman (1994) and Pagel (1993) have estimated that
$\sigma _{syst.} \sim 0.005$, Copi, Schramm, \& Turner (1994) have
suggested
much
larger uncertainties
$\sim$ 0.016. The resolution of this potential crisis will require  careful
analyses of the data from the  most metal-poor extragalactic H~{\eightrm
II} regions
(e.g.,
Skillman \& Kennicutt 1993).

\noindent (b) Is the Predicted D and/or $^3$He  Yield Too Low?

A second possible crisis involves recent data from a QSO absorption system
where --
possibly -- deuterium has been observed (Songaila et al.\ 1994, Carswell et
al.\
1994). The caveat is that  there is no way to distinguish an individual
absorption system due to D from one due to a hydrogen ``interloper"
(Steigman 1994b).
If, however, the observed absorption features are, indeed, due to
deuterium, the
derived abundance is surprisingly high (D/H $\sim 2 \times 10 ^{-4}$),
nearly an
order of magnitude higher than those derived from interstellar (Linsky et
al.\ 1993)
and/or solar system (Geiss 1993) data. Even so, this is not a crisis for
BBN (Steigman
1994a,b) since such a high primordial D abundance forces us to a low value
of
$\eta$, reducing the (possibly problematic) predicted yield of \hel4.
Indeed, for
  (D/H)$_{BBN} \approx 2 \times 10^{-4},~{\rm we ~obtain}~ \eta _{10}
\approx 1.5
{}~{\rm
and}~ Y_{BBN}
\approx 0.23 ~[{\rm as~ well~as} ~ (^7{\rm Li/H})_{BBN} \approx 2 \times
10^{-10}$,
which  is entirely
consistent with the observational data].  This lower value of $\eta$ leads
to a
lower $\Omega _{BBN}$ which remains consistent with  ($\ge)~ \Omega
_{LUM}$,
but which
reinforces the evidence for non-baryonic dark matter $(\Omega _{BBN} <
\Omega
_{DYN})$. However, such a high primordial D abundance does challenge models
of
stellar and galactic chemical evolution since it would require that $\sim $
90 \% of
the pregalactic deuterium would have had to be destroyed in stars while
avoiding the
overproduction of
$^3$He  (Steigman 1994b). Anticipated (and rumored!) additional data on
possible D in other QSO absorption systems from Keck and the HST will help
resolve --
or sharpen -- this potential crisis.

It is, however, the third crisis -- the x-ray cluster baryon crisis -- we
wish to
consider more fully here.

\noindent 3. THE X-RAY CLUSTER CRISIS

Let's suppose for the moment that rich clusters of galaxies provide a
``fair sample"
of baryonic vs.\ non-baryonic mass in the Universe. Then the baryon
fraction of the
total mass in a cluster should be the same as the universal baryon fraction:
$$
 f_B \equiv  (M_B/M_{TOT})_{clusters} =\Omega _B/\Omega .
\eqno (4)$$

Many rich clusters are x-ray sources, the emission being due to a hot
intracluster gas (of baryons and electrons!). For such clusters (Loewenstein
1994) the x-ray emission provides information on the mass in hot gas (from
the observed angular distribution of temperature and surface brightness) and
on the total mass  (from imposing a requirement of thermal
hydrostatic equilibrium on the gas). In addition to the hot gas, baryons
are, of
course, to be found in the cluster galaxies. However, for most observed rich
clusters the hot gas apparently dominates the baryon budget. In any case,
ignoring
the baryonic contribution by galaxies leads to a bound on
$f_B ~(f_{HG} ~\le f_B)$, so that
$$
\Omega ~\le \Omega _B/f_{HG}.
\eqno(5)$$

Recent measurements show $f_{HG}$ to be rather large, $\sim$ 0.2. The value
of
$f_{HG}$ inferred for an x-ray cluster depends on the distance to the
cluster and,
hence, on
 $H_0 ~(f_{HG} \propto H{_0 ^{-3/2}})$. If we define $f_{50}$ by
$$
f_{HG} \equiv f_{50}h_{50} ^{-3/2} \eqno (6)
$$
and use  (2) for $\Omega _B$, we may write
$$
 \Omega ~\le 0.3 h{_{50} ^{-1/2}} \bigg ( {0.20 \over f_{50}} \bigg )
\bigg (
{\eta _{10} \over 4.0} \bigg ).
\eqno (7)$$

{}From BBN we have concluded that $ \eta _{10} \le 4$, so that, for $H_0 \ge
40,
\Omega \le (1/3) (0.2/f_{50}) $ and, unless $f_{50}$ is $\ll 0.2$, we are
led
to
conclude that $\Omega < 1$, rejecting the popular Einstein-de Sitter
cosmology.
This, in a nutshell, is the x-ray cluster crisis.

White et al.\ (1993, ``WNEF"), in an important paper, analyzed the data for
the Coma
cluster and derived
$f_{50} {\rm (Coma)}   \approx 0.14 \pm 0.04$. For $f_{50} \ge 0.10 ~{\rm
and}~ \eta _{10} \le 4.0, \Omega =1 ~ {\rm would ~require}~ H_0 \le 18~
{\rm km~s^{-1}Mpc} ^{-1}$! Alternatively, for $H_0  \ge 40$, we are led
to infer $\Omega \le 2/3$.

How serious is this ``crisis"? From  (7) it would appear that if
$H_0$ and
$f_{50} {\rm (Coma)}$ are near the lower ends of their ranges $(H_{0}
\approx 40,~
f_{50}
\approx 0.1)~ \Omega =1$ could be recovered for $\eta _{10} \approx 6$.
BBN theorists might be bullied into accepting this (Copi et al.\ 1994).
However, the
x-ray  data suggest that the crisis is, in fact, much worse. In their
analysis of
Coma, WNEF were ``conservative" in the sense of having chosen the
largest of several estimates for the {\it total} mass. For example, had
they used
their mass estimate based on the assumption that the optical light traces
the mass,
the WNEF result would have been $f_{50} {\rm (Coma)} \approx 0.23$.

There is, indeed, accumulating support for larger values of $f_{50}$. In
their
recent analysis of the Coma data, Fusco-Femiano \& Hughes (1994) find
(within the
Abell radius of $ 3h{_{50} ^{-1}} ~{\rm Mpc}) ~ f_{50} \approx 0.27$. For
the
cluster Abell
478, White et al.\ (1994) find (within a radius of $2.3 h{_{50} ^{-1}}~{\rm
Mpc})
{}~f_{50} \ge 0.28$.

Even earlier, Fabian (1991), in his analysis of the core of the Shapley
supercluster, concluded that $f_{50} > 0.18$. B\"ohringer (1994) concludes
from
ROSAT data that three massive clusters (Coma, Perseus and A2256) are very
similar to one another and have $0.12 < f_{50} < 0.45$.  (The range here
arises
from uncertainty in the modelling, because ROSAT does not give good
information
on the temperature distributions.) Higher baryonic fractions receive further
support from recent reanalyses of extant x-ray cluster data (Durret et al.\
1994,
White \& Fabian 1994). In both recent studies there is a clear trend of
$f_{50}$ increasing with cluster size. The data of White \& Fabian (1994)
for
19 clusters suggest that $f_{50} \approx 0.12  R{_{50} ^{0.6}}$ where
$R_{50}~
{\rm in~ Mpc}$, for each cluster, is the largest radius   observed
(for $H_0
= 50$). Evaluating this at the  Abell radius of Coma ($R_{50} = 3 ~$Mpc)
gives
$f_{50}
\approx 0.23$.
Finally, data from the ASCA satellite, which gives good information on the
temperature distributions, imply that for three rich regular clusters
(A496, A1795
and A2199), $f_{50} \approx 0.20$ within $\sim 0.75 h{_{50} ^{-1}} ~$Mpc,
rising to
$f_{50} \approx 0.23 ~{\rm within}~ \sim 2 h{_{50} ^{-1}} ~$Mpc (Mushotzky
1995).
The baryons in the cluster galaxies, which we have neglected, would
increase any of
these $f_{50}$ values by 0.01--0.04 (WNEF, Mushotzky 1995).

 Thus, it would appear that
the x-ray cluster baryon crisis is worse, by a factor of
$\sim$ 2, than that identified for Coma by WNEF.  For $f_{50} \ge
0.2$, we are driven to
$$
 \Omega h{_{50} ^{1/2}} ~\le 0.3 ~(\eta _{10} /4).
\eqno (8)$$
{}From this perspective it seems far less likely that an overly restrictive
BBN  limit
$\eta _{10} \le 4$ is the ``culprit" since, for $H_0 \ge 40$ (80; Freedman
et al.\
1994), $\eta _{10} \ge 12 (17)$ would be required to ``save" $\Omega = 1$.
For such
large values of $\eta$ the BBN yields of D   and $^3$He are far below their
observed
values while, far more seriously, the abundances of \hel4 and \li7 exceed
those
observed.
\vfil\eject

\noindent 4. MITIGATING THE CRISIS

Let's search for ways to mitigate this crisis. One or more of the following
possibilities might contribute to a solution. We discuss several of these
in more
detail elsewhere (Felten \& Steigman 1994).

\noindent (a) Fair Sample?

The fair-sample hypothesis is an obvious point of attack. WNEF studied this
by
defining a ``baryon enhancement factor" $\Upsilon$, replacing
  (4)  with
$$
f_B \equiv (M_B/M_{TOT})_{cluster} = \Upsilon \Omega _B /\Omega.
\eqno (9) $$
We can imagine that clustering processes could produce $\Upsilon > 1 ~{\rm
or}~ <
1$, in individual clusters or in all clusters. From  (8), a value
$\Upsilon
\approx 3 (4) ~{\rm for}~ H_0 = 50 (80)$ would dispel the crisis. However,
modern
simulations by WNEF with $\Omega = 1$ and cold dark matter (CDM) produce
$\Upsilon
\le 1.4$, and in fact $\Upsilon$ tends to drop below unity at large cluster
radii.
Cen \& Ostriker (1993) found $\Upsilon \approx 2/3$ for Coma-like clusters
in their
simulations. This is a generic property and is likely to persist in other
simulations with $\Omega < 1$ and/or non-zero cosmological constant
$\Lambda$. The
reason is that gas can support itself partially against collapse through
pressure,
turbulence and shocks, but cold dark matter cannot. On smaller scales $(\le
100~
$kpc, the scales of galaxies), where the gas density can rise high enough
for
cooling to become important, the gas can indeed concentrate, but this does
not
happen on scales as large as 1 Mpc.

Larger
values of $\Upsilon$ can be obtained by using cold plus hot dark matter
(CHDM)
simulations, still with $\Omega = 1$ (Bryan et al.\ 1994, Primack et al.\
1994). Hot
dark matter can stay out of the clusters. The trouble is that HDM cannot
dominate;
otherwise the right kind
of structure is not produced at the right times. The maximum $\Omega
_{HDM}$ is
about 0.2 -- 0.3. This is so small that even if all the HDM avoids the
clusters,
$\Upsilon$ cannot be large. People doing these simulations tend to press
BBN theory
rather hard. For example, Bryan et al.\ (1994) took $\eta _{10} \approx 7$,
which
may be  unacceptable for BBN. Even with $\eta _{10}$ this large, Klypin
(1995)
reports that the maximum $f_{50}$ for clusters obtained in the simulations
is about
0.15. This is still too small to match the recent observations.

All the simulations above use Gaussian fluctuations as seeds to grow
structure.
  Schramm (1994) and   White (1995) have suggested that topological
seeds
(strings, etc.) in a HDM-dominated Universe ($\Omega _{HDM} \approx 0.7 -
0.8$,
perhaps) might grow the right kind of structure and also produce large
$\Upsilon$.
This is a possibility, but we have not seen any such simulations.

\noindent (b) $H_0 = 30$ ?

Bartlett et al.\ (1994) suggest that a resolution is to be found in a very
small
Hubble
constant; their favored value is $H_0 = 30$. From Figure 1 it is clear that
$\Omega
_{BBN}$ can reach the value of $f_{50} \approx 0.2$, doing away with the
need for a
baryon enhancement factor $\Upsilon$, if $H_0$ takes some small value.
Observationally, we think that $H_0 = 30$ is  a counsel of despair (cf.
Freedman et
al.\ 1994).  Furthermore, $H_0 = 30$ is not small enough. From   (8), for
$H_0 =
30$, we obtain
$\Omega = 1$ only if
$\eta _{10}
\approx 10$, which is untenable.

\noindent (c) Magnetic Fields, Turbulence, Clumping

Lensing determinations of total masses of clusters give, in a few cases,
masses higher
by factors 2--2.5 than the total masses obtained in the x-ray analyses
(Miralda-Escud\'e
\& Babul 1994, Bartelmann \& Narayan 1995). If the x-ray total masses are
too small
by this factor, that would go more than halfway toward resolving the
crisis. This
has given rise to the suggestion that the x-ray assumption of thermal
hydrostatic
equilibrium is wrong. If magnetic pressure, for example, is in
equipartition with
thermal pressure in the hot gas, we can add this to the hydrostatic
equation. The
mass of hot gas remains the same, but the total (gravitational) mass
inferred
doubles, so that $f_{50}$ is reduced by a factor 2. If turbulence is also in
equipartition, the total mass triples.

The fields needed are large. Loeb \& Mao (1994), requiring equipartition,
predict
specifically $B \sim 50 ~\mu G$ in the cluster A2218, where there are no
measurements
of $B$. This may be important in a few clusters, but we doubt that it can
provide a
general solution to the crisis. The Coma cluster field is not nearly large
enough
(Kim et al.\ 1990), and cluster fields generally are thought to be $\sim 1-
2~
\mu G$
(Kronberg 1994).

The lensing determinations have their own problems, and need refinements in
some
cases (Bartelmann \& Narayan 1995). Cluster masses have also been
determined by the
familiar use of galaxy velocity dispersions (``virial-theorem masses").
These are
subject to factor-of-2 systematic errors arising from models of orbit
shapes and of
the radial mass distributions, but (within errors) they agree with the x-
ray masses
(WNEF, Mushotzky 1995). The galaxies (unlike the gas) cannot be supported by
$B$-fields and turbulence, so the agreement between x-ray and virial-
theorem
masses
puts some limit on the importance of fields and turbulence.

There is also a related suggestion that the gas may be strongly clumped.
Since the
emissivity of the gas goes as $n^2$, clumping increases the pressure but
reduces
the mass of gas required (WNEF) and reduces $f_{50}$. One might then expect
to see a
superposition of many different temperatures (in different clouds) in the x-
ray
data. There is no evidence of this. We also have to question the stability
of such a
clumped system. If some kind of equilibrium is maintained, then something,
e.g., a
large $B$-field $(> 50~ \mu G$ in A2218!), must supply pressure in the
voids
yet
remain outside the clouds---astrophysically a bizarre situation. If there
is no
equilibrium, we could think of the clouds as moving freely on orbits. In
either
case, the clouds might dissipate rapidly. And in either case, the
calculation of
total mass from x-ray data would be erroneous, and the existing agreement
between
the x-ray and virial-theorem masses would be an accident (WNEF).

\vfil\eject

\noindent 5. COSMOLOGY AND DICKE COINCIDENCES

We come now to the possibility that the ``standard", i.e.,\ Einstein-de
Sitter,
cosmology should be abandoned. To comment on this, we write the Friedmann
equation:
$$
1 =  {8 \pi G \rho _{M0} \over 3 H^2}  \Bigg ( {a_0 \over a}
\Bigg )^3
-  {c^2 K_0 \over H^2} \Bigg ( {a_0 \over a} \Bigg )^2 +
{\Lambda \over
3 H^2 }~ ,
\eqno (10)$$
where $\rho _M$ is mass density, $a(t)$ is the scale factor, $K$ is the
Gaussian
curvature of three-space, subscripts 0 denote present values, and $H(t)$ is
the
Hubble function $(\dot a /a)$. (The dependence of the first term above on
$a$
steepens to  the Lema\^ itre form $(a_0/a)^4$ in the early
radiation-dominated universe.) The three terms have familiar shorthand forms
(Peebles 1993, p.\  100):
$$
1 = \Omega _M + \Omega _K + \Omega _{\Lambda} .
\eqno (11)$$
($\Omega _{M0}$ is the quantity we have called $\Omega$ earlier.
We
switch notation to discuss the three terms more easily.)

Many cosmologists are attached to the Einstein-de Sitter model, in which
the last
two terms are negligible at present. One still hears often (e.g., Fujii \&
Nishioka
1991) the claim that these two terms must be negligible to avoid the curse
of the
``Dicke coincidences" (Dicke \& Peebles 1979). If Einstein's gravity is
correct, we
know astronomically that at present $\Omega _{M0}$ is not terribly small
$(\ll
10^{-2}$) and the other two terms are not terribly large $(>> > 1)$.
Because these two
terms shrink relative to $\Omega _M$ as we look back into the past $(a \to
0)$, it
follows that $\Omega _M$ was extremely close to unity at early epochs
(e.g., at
epoch ``b", the epoch of BBN). This is the ``flatness problem". Many
believe that
the only way to explain this fantastic coincidence is to postulate that
$\Omega
_{Mb}$ was (and $\Omega _{M0}$ remains) exactly unity. This would agree
with the
idea that $\Omega _K$ was driven to zero by inflation, and that $\Omega
_{\Lambda}$
is zero because of some unknown physics.

But in fact, $\Omega _{Mb}$ {\it must} be close to unity in {\it any}
cognizable
(``anthropic") Universe (Barrow \& Tipler 1986, pp.\ 408 ff.; Peebles 1993,
pp.\ 364
ff.). Consider, for a moment, only the familiar $\Lambda = 0$ models. If
$\Omega
_{Mb}$ is appreciably below unity, the early Universe goes into a linear
expansion in
a few doubling times (i.e., a few minutes) and according to present ideas of
cosmogony it could never form structure. If $\Omega _{Mb}$ exceeds unity,
the
Universe recollapses, again in a few minutes. All of these models are
hostile to
life. Thus the anthropic principle gives a solution to the flatness problem
without
the postulate that $\Omega _{Kb}$ is exactly zero.

Sizable $\Omega _{\Lambda b}$, it seems, would not add anthropic
possibilities. Such a
model could coast at a hot dense epoch, but upon emerging, it would go into
fast
recollapse or accelerated expansion. [The ``flow" of the $\Omega$ terms has
been
studied (e.g., Madsen \& Ellis 1988, Ehlers \& Rindler 1989, Cho \&
Kantowski 1994).]

Anthropic arguments like these discomfit some physicists, probably be\-
cause
advocates
sometimes bring in teleological and even theological baggage. Kolb \&
Turner (1990),
writing in their entertaining style, call the anthropic principle ``lame"
(p.\ 269),
but then relent and concede (p.\ 315) that it has ``some rational basis".
We believe
that anthropic arguments can be quite powerful and are certainly worthy to
answer
``coincidence" arguments.

Consider now the present (epoch ``0") values of the terms in  (11).
We
grant, because of the argument above, that $\Omega _{K0}$ and $\Omega
_{\Lambda 0}$
need not be exactly zero, but we note that all three terms have different
$a$-dependences. Is it then a big coincidence if more than one of the terms
are
$\sim 1$ at present? Consider $\Omega _{\Lambda 0}$. The physically
``natural"
value for $\Omega _{\Lambda 0}$ (Weinberg 1989; Carroll, Press, \& Turner
1992) seems
to be $\sim \pm 10 ^{118}$! We could imagine that there is a natural
distribution of
possible magnitudes with this as mean. But anthropically $\Omega _{\Lambda
0}$ is
limited roughly to
$$
 -10 < \Omega _{\Lambda 0} < 100 \Omega _{M0}
\eqno (12)$$
(Weinberg 1987, 1989; cf. Barrow \& Tipler 1986, inequality 6.131). The
argument is
essentially the same as that given above. $\Omega _{\Lambda 0} < -10$ makes
the
Universe recollapse too quickly; $\Omega _{\Lambda 0} > 100  \Omega _{M0}$
prevents
the growth of condensations (structure). The upper limit has been derived
rigorously
only for flat models, and there are some errors in published derivations.
The
numbers may change somewhat, but rough numbers are adequate to make our
point, which
is this: Once we realize that $\Omega _{\Lambda 0}$ is not free to roam up
to $10
^{118}$, but is constrained below 100 or so by the condition that
intelligent life
exists, the ``coincidence" involved if $\Omega _{\Lambda 0} \sim 1$ is
no longer
so impressive. Weinberg (1989) suggests that the true value should be near
the upper
limit in  (12) ``because there is no anthropic reason for it to
be any
smaller". We wonder whether the ``natural" probability distribution of
$\Omega
_{\Lambda 0}$ is established well enough to justify that statement. We
think that an
astronomically admissible and cosmogonically significant value (say, 0.2--
0.7) may
be plausible.

This is already acknowledged, for many simulations have been done with
$\Omega
_{\Lambda 0} > 0$. Should we also consider non-zero $\Omega _{K0}$?
Inflation
requires $\Omega _{K0} = 0$; models in which inflation stops short of
flattening the
Universe encounter difficulties (Kashlinsky, Tkachev \& Frieman 1994).
Inflation
gives a nice solution to the horizon problem, but its predictive power has
not been
great. What if we abandon the inflationary model? Should we still maintain
that
$\Omega _{K0} = 0$? If all three terms in  (11) are $\sim 1$, is
that an
unacceptable ``double coincidence"?

This is a bit murky, but several points should be made:

(1) $\Omega _{K0}$ is not really free, being determined by   (11).

(2) Nevertheless there are two free variables in   (11), because
$\Omega
_{M0}$ is free. In our Universe we know its value very roughly, but in other
cognizable Universes its value might be very different. Note that it
affects the
upper limit in  (12).

(3) In principle there are anthropic limits on the ratios of $\Omega _{K0}$
to the
other terms (Carter 1974, Carr \& Rees 1979), analogous to  (12).
$\Omega
_{K0}$ is not free to roam to extreme values.

We have not seen a derivation of joint anthropic constraints on the terms in
 (11) when all three are allowed to be non-zero, but we can guess at
the
results. We believe that the case $\Omega _{K0} = 0$ deserves some
preference
because of inflation theory. In alternative models with  $\Omega _{K0}
\not= 0$,
we would guess that the case  $\Omega _{\Lambda 0} = 0$ is more likely than
the
case of all
three terms $\sim 1$, but not by a large factor when anthropic
constraints are
considered. We cannot conclude safely from coincidence arguments that any
of the terms
in   (11) is negligible. Anthropic constraints are numerous and
powerful
(Carter 1974, Carr \& Rees 1979, Barrow \& Tipler 1986); we have mentioned
only the
simplest ones.

Perhaps some of these ideas for mitigating the x-ray cluster crisis will be
productive. Perhaps simulations with topological seeds and hot dark matter
can
produce the right kind of structure and achieve a baryon enhancement factor
$\Upsilon \approx 3 - 4$, so that $\Omega _{M0} =1$ can be maintained. If
not,
open-universe models will necessarily gain in popularity (Cen \& Ostriker
1993).

\bigskip
\noindent ACKNOWLEDGEMENTS

We wish to thank Sid Bludman, Richard Ellis, Andy Fabian, Sasha Kashlinsky,
Rich
Mushotzky,  Martin Rees and
Simon White for advice and discussion.  J. E. F. is grateful to NASA for
partial
support under RTOP No. 188-44-53-05, and wishes to thank the Aspen Center
for
Physics and all the participants in the Aspen 1994 Workshops on
Gravitational
Clustering and on Clusters of Galaxies.  The work of G. S. at OSU is
supported by
DOE grant DE-FG02-94ER-40832.  Some of this work was done when G. S. was an
Overseas Fellow at Churchill College, Cambridge and a Visiting Fellow at the
Institute of Astronomy and he thanks them for hospitality.
G. S. thanks the organizers of  the Gamow Seminar for their kind
hospitality and
Roger Chevalier for assistance in preparing this manuscript for the
Proceedings.
 \vfil\eject

\noindent REFERENCES

\noindent  Bartelmann, M., \& Narayan, R.:
 1995, in ``Dark Matter" (Proc. 5th  Annual
Astro-\break\indent physics Conference in Maryland, Oct. 10-12, 1994), ed.
S. S. Holt
\&  C. L.\break\indent
Bennett (New York: American Institute of Physics), in press

\noindent Barrow, J. D., \& Tipler, F. J.: 1986, ``The Anthropic
Cosmological
Principle"
(Ox-\break\indent ford: Oxford University Press)

\noindent Bartlett, J. G., Blanchard, A., Silk, J., \& Turner, M. S.: 1994,
{\it
Science},
submitted

\noindent Boesgaard, A.M., \& Steigman, G.: 1985, {\it ARA\&A},
{\bf 23}, 319

\noindent B\"ohringer, H.: 1994, preprint

\noindent Bryan, G. L.,  Klypin, A., Loken, C., Norman, M. L., \& Burns, J.
O.:
1994,
{\it APJLett},\break\indent {\bf 437}, L5

\noindent  Carr, B. J., \& Rees, M. J.: 1979, {\it Nature}, {\bf  278}, 605

\noindent  Carroll, S. M., Press, W. H., \& Turner, E. L.: 1992, {\it
ARA\&A},
{\bf 30},
499

\noindent  Carswell, R. F, Rauch, M., Weymann, R. J., Cooke, A. J., \&
Webb, J. K.:
1994,
\break\indent{\it
MNRAS\/}, {\bf 268}, L1

\noindent  Carter, B.: 1974, in ``Confrontation of Cosmological Theories
with
Observational\break\indent
Data" (IAU Symposium 63, Cracow, September 10-12, 1973), ed. M. S. Longair
\break\indent (Dordrecht:
Reidel), 291

\noindent  Cen, R., \& Ostriker, J. P.: 1993, {\it ApJ}, {\bf 417}, 404

\noindent  Cho, H. T., \& Kantowski, R.: 1994, {\it Phys. Rev.}, {\bf D50},
6144

\noindent  Copi, C. J., Schramm, D. N., \& Turner, M. S.: 1994, {\it
Science,}
submitted

\noindent  Dearborn, D. S. P., Schramm, D. N., \& Steigman, G.: 1986, {\it
Ap.\
J.} {\bf
302}, 35

\noindent  Dicke, R. H., \& Peebles, P. J. E.: 1979, in ``General
Relativity: An
Einstein
Cente-\break\indent nary Survey", ed. S. W. Hawking \& W. Israel (Cambridge:
Cambridge  Univer-\break\indent sity
Press), 504

 \noindent Durret, F., Gerbal, D., Lachi\`eze-Rey, M., Lima-Neto, G., \&
Sadat, R.:
1994, {\it A\&A},\break\indent {\bf 287}, 733

 \noindent Efstathiou, G., Ellis, R. S., \& Peterson, B. A.: 1988, {\it
MNRAS},
{\bf 232},
431

\noindent  Ehlers, J., \& Rindler, W.: 1989, {\it MNRAS}, {\bf 238}, 503

\noindent  Fabian, A. C.: 1991, {\it MNRAS}, {\bf 253}, 29P

\noindent  Felten, J. E. \& Steigman, G.: 1994, in preparation

\noindent  Freedman, W. L., et al.: 1994, {\it Nature}, {\bf 371}, 757

\noindent  Fujii, Y., \& Nishioka, T.: 1991, {\it Phys. Lett.}, {\bf B254},
347

\noindent  Fusco-Femiano, R., \& Hughes, J. P.: 1994, {\it ApJ}, {\bf 429},
545

\noindent  Geiss, J.: 1993, in ``Origin and Evolution of the Elements",
ed. N.
Prantzos,
\break\indent E. Vangioni-Flam \& M. Cass\'e,  (Cambridge: Cambridge
University
Press), 89

 \noindent  Kashlinsky, A., Tkachev, I.I., \& Frieman, J.: 1994, {\it Phys.
Rev.
Lett.}, {\bf 73}, 1582

 \noindent Kim, K.-T., Kronberg, P. P., Dewdney, P. E., \& Landecker, T.
L.: 1990,
{\it
ApJ}, {\bf 355}, \break\indent 29

 \noindent Klypin, A.: 1995, in ``Dark Matter" (Proc. 5th Annual
Astrophysics
Confe-\break\indent rence in
Maryland, Oct. 10-12, 1994), ed. S. S. Holt \& C. L. Bennett (New York:
\break\indent
American
Institute of Physics), in press

\noindent  Kolb, E. W., \& Turner, M. S.: 1990, ``The Early Universe"
(Redwood
City, CA:\break\indent
Addison-Wesley)

 \noindent Kronberg, P. P.: 1994, {\it Rep. Prog. Phys.}, {\bf 57}, 325

\noindent Linsky, J. L., et al.: 1993, {\it ApJ\/}, {\bf 402}, 694

 \noindent Loeb, A., \& Mao, S.: 1994, {\it ApJ}, {\bf 435}, L109

 \noindent Loewenstein, M.: 1994, {\it ApJ}, {\bf 431}, 91

\noindent  Madsen, M. S., \& Ellis, G. F. R.: 1988, {\it MNRAS}, {\bf 234},
67

 \noindent Mather, J. C., et al.: 1994, {\it ApJ}, {\bf 420}, 439

 \noindent Miralda-Escud\'e, J., \& Babul, A.: 1994, {\it ApJ}, submitted

 \noindent Mushotzky, R. F.: 1995, in ``Dark Matter" (Proc. 5th Annual
Astrophysics
Confer\break\indent ence in Maryland, Oct. 10-12, 1994), ed. S. S. Holt \&
C. L.
Bennett  (New
York:\break\indent  American Institute of Physics), in press

\noindent  Olive, K. A., \& Steigman, G.: 1994, {\it ApJS\/}, in press

\noindent Pagel, B. E. J.: 1993, {\it Proc. Nat. Acad. Sci.}, {\bf 90}, 4789

\noindent  Pagel, B. E. J., Simonson, E. A., Terlevich, R. J., \& Edmunds,
M. G.:
1992,
\break\indent {\it
MNRAS\/}, {\bf 255}, 325

 \noindent Peebles, P. J. E.: 1993, ``Principles of Physical Cosmology"
(Princeton:
Princeton \break\indent University Press)

 \noindent Primack, J. R., Holtzman, J., Klypin, A., \& Caldwell, D. O.:
1994,
{\it Phys.
Rev. \break\indent Lett.}, submitted

 \noindent Schramm, D. N.: 1994, private communication

\noindent Skillman, E. D., \& Kennicutt, R. C., Jr.: 1993, {\it ApJ}, {\bf
411}, 655

\vfil\eject
 \noindent Skillman, E. D., Terlevich, R. J., Terlevich, E., Kennicutt, R.
C., Jr., \&
Garnett, \break\indent D. R.: 1993, in ``Texas/PASCOS '92: Relativistic
Astrophysics
and Particle\break\indent Cosmology", ed. C. W. Akerlof \& M. A. Srednicki
({\it Ann.
N. Y. Acad. Sci.}, {\bf 688})\break\indent (New York: New York Academy of
Sciences),
739

 \noindent Songaila, A., Cowie, L. L., Hogan, C. J., \& Rugers, M.: 1994,
{\it
Nature\ },  {\bf 368},
599

 \noindent Steigman, G.: 1994a,  preprint OSU-TA-22/94. To appear in the
proceedings
of the \break\indent ESO/EIPC Workshop on ``The Light Element Abundances",
ed.\ P.\
Crane\break\indent (Berlin: Springer)

 \noindent \hbox{\vrule width 0.35in height 0pt depth .5pt \ .: 1994b, {\it
MNRAS\/},
{\bf 269}, L53}

\noindent   Steigman, G., \& Tosi, M.: 1992, {\it ApJ\/} {\bf 401}, 150

\noindent   \hbox{\vrule width 0.35in height 0pt depth .5pt \ .: 1994, {\it
ApJ},
submitted}

\noindent  Walker, T. P., Steigman, G., Schramm, D. N., Olive, K. A., \&
Kang, H.-S.:
1991,\break\indent  {\it
ApJ\/}, {\bf 376}, 51 (WSSOK)

 \noindent Weinberg, S.: 1987, {\it Phys. Rev. Lett.}, {\bf 59}, 2607

 \noindent \hbox{\vrule width 0.35in height 0pt depth .5pt \ .: 1989, {\it
Rev. Mod
Phys.}, {\bf 61}, 1}

 \noindent White, D. A., \& Fabian, A. C.: 1994, {\it MNRAS\/}, in press

\noindent  White, D. A., Fabian, A. C., Allen, S. W., Edge, A. C.,
Crawford, C. S.,
Johnstone,\break\indent R. M., Stewart, G. C., \& Voges, W.: 1994, {\it
MNRAS\/},
{\bf 269}, 589

 \noindent White, S. D. M.: 1995, in ``Dark Matter" (Proc. 5th Annual
Astrophysics
Confe-\break\indent rence in Maryland, Oct. 10-12, 1994), ed. S. S. Holt \&
C. L.
Bennett  (New\break\indent York:
American Institute of Physics), in press

\noindent  White, S. D. M., Navarro, J.
F., Evrard, A. E.,
\& Frenk, C. S.: 1993, {\it Nature\/} {\bf 366},\break\indent 429 (WNEF)

\noindent  Yang, J., Turner, M. S., Steigman, G., Schramm, D. N., \& Olive,
K. A.:
1984,  {\it
ApJ\/},\break\indent {\bf 281}, 493

 \vfil\eject

\noindent FIGURE CAPTION

\noindent
FIG. 1 --- The range of total baryonic mass density $\Omega _{BBN}$ allowed
by BBN
theory and element abundances (for $2.8 \le \eta _{10} \le 4.0$), shown as a
function of assumed Hubble constant $H_0 ({\rm km~s} ^{-1} {\rm Mpc} ^{-
1}$). We also
show a generous estimate for the mass density $\Omega _{LUM}$ due to
baryons in the
luminous parts of galaxies, and a lower limit to the total gravitating mass
density
$\Omega _{DYN}$ implied by large-scale dynamical studies.

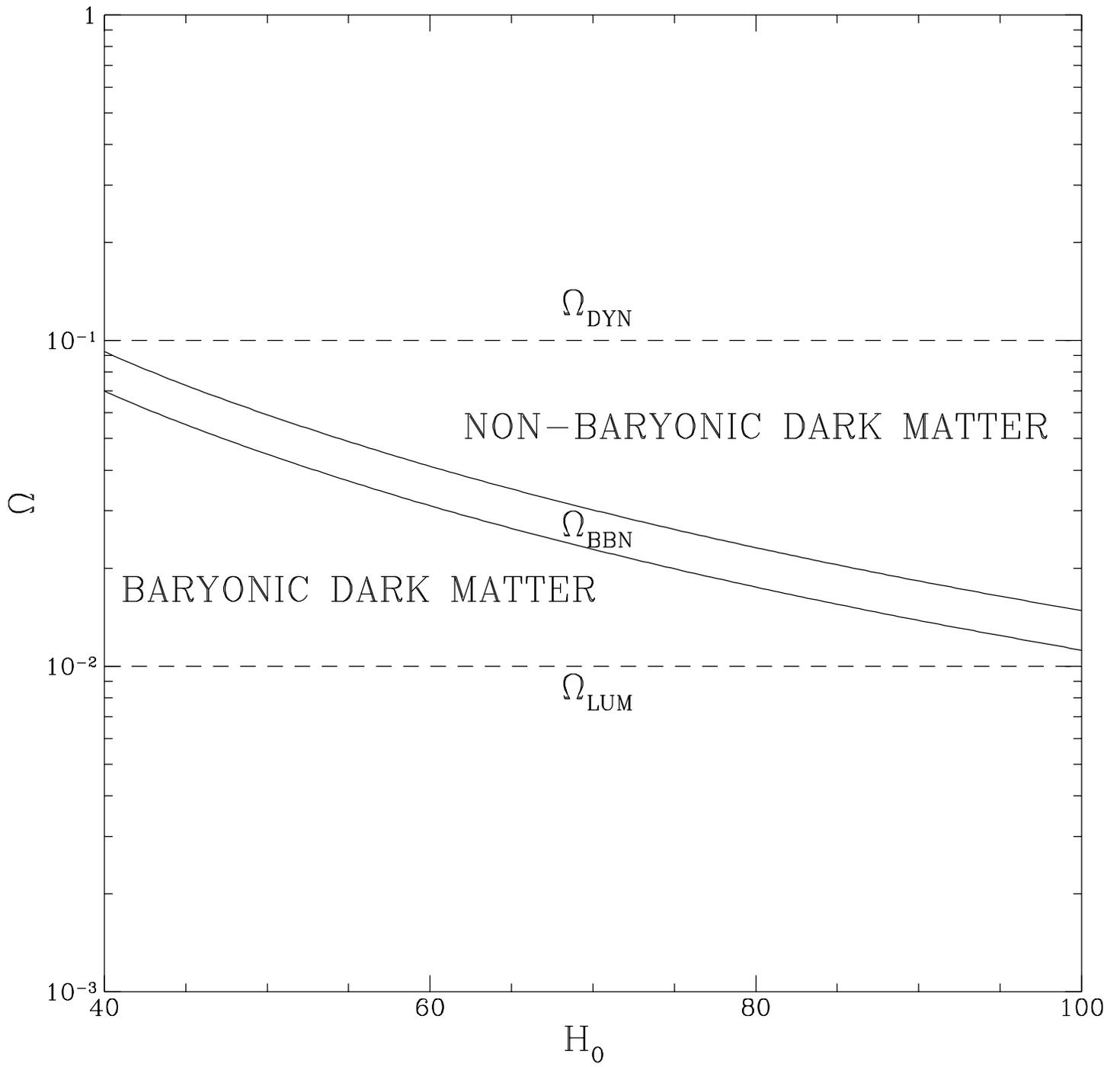
\end{document}